\documentclass[twocolumn,secnumarabic,amssymb, nobibnotes, aps, prl]{revtex4-1}

\usepackage{epsfig}
\usepackage{graphicx}
\usepackage{dcolumn}
\usepackage{bm}
\usepackage[utf8]{inputenc}

\begin{document}

\title{Selective spin and nematic excitations using polarized pump pulses}

\author{Ganesh Adhikary$^{1\dag}$}
\altaffiliation{Present address: Faculty of Science and Technology, ICFAI University, Tripura 799210, India.}
\author{Tanusree Saha$^1$}
\author{Primo\v{z} Rebernik Ribi\v{c}$^{1,2}$}
\author{Matija Stupar$^1$}
\author{Barbara Ressel$^1$}
\author{Jurij Urban\v{c}i\v{c}$^1$}
\author{Giovanni De Ninno$^{1,2\dag}$}
\author{A. Thamizhavel$^3$}
\author{Kalobaran Maiti$^{3}$}
\altaffiliation{Corresponding authors: ganesh.adhikary@gmail.com, giovanni.de.ninno@ung.si, kbmaiti@tifr.res.in}

\affiliation{$^1$Laboratory of Quantum Optics, University of Nova Gorica, 5001 Nova Gorica, Slovenia.}
\affiliation{$^2$Elettra-Sincrotrone Trieste, Area Science Park, 34149 Trieste, Italy.}
\affiliation{$^3$Department of Condensed Matter Physics and Materials Science, Tata Institute of Fundamental Research, Homi Bhabha Road,
Colaba, Mumbai - 400 005, INDIA.}

\begin{abstract}
Quantum materials display exotic behaviours related to the interplay between temperature-driven phase transitions. Here, we study electron dynamics in one such material, CaFe$_2$As$_2$, a parent Fe-based superconductor, employing time and angle-resolved photoemission spectroscopy. CaFe$_2$As$_2$ exhibits concomitant transition to spin density wave state and nematic order below 170 K. We discover that magnetic excitations can be induced selectively, by using polarized pump pulses, significantly before the destruction of nematic order. More specifically, we observe that $s$-polarized light enhances electron temperature by several hundreds of degrees at a time scale of about 200 fs, while the temperature of the electrons participating in magnetic order can be enhanced by similar amount at a much faster time scale (50 fs) using $p$-polarized pump pulse. These results provide a pathway to achieve selective electron heating, which is not possible with other methods, as well as to disentangle phase transitions in quantum materials.
\end{abstract}

\date{\today}


\maketitle

It is believed that exoticity in quantum materials arises due to complex interplay between spin, charge and orbital degrees of freedom. This makes the research on these materials challenging. A recent addition to this group are the Fe-based compounds \cite{Kamihara1,Kamihara2,iron-pnictide1,iron-pnictide2}, often exhibiting complex phase diagrams, involving magnetic order, superconductivity and structural transitions. The coexistence of mutually exclusive, magnetic order and superconductivity is one of the many outstanding puzzles observed in these materials, questioning existing theoretical models \cite{EuFe2As2}. Numerous investigations allowed scientists to shed light into the fundamental properties of Fe-based systems: the parent compounds exhibit spin density wave (SDW) state at low temperature, accompanied by a structural transition from tetragonal to orthorhombic phase \cite{iron-pnictide1,iron-pnictide2}. Since multiple bands contribute in the formation of Fermi surfaces, it is difficult to disentangle the role of different orbitals/electronic states in determining the electronic properties of the material.

\begin{figure}
\includegraphics[width = 0.95\linewidth]{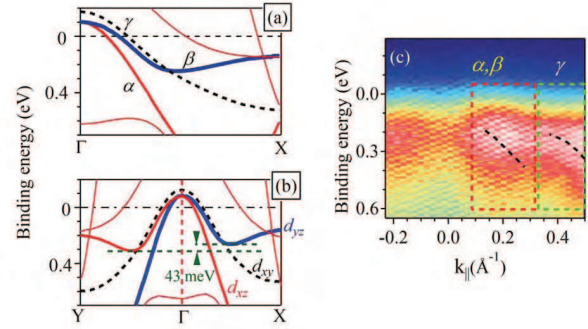}
\caption{(color online) Calculated band structure (a) in tetragonal phase along $\Gamma X$ and (b) in orthorhombic phase along $\Gamma X$ and $\Gamma Y$. In the orthorhombic phase, the degeneracy of $d_{xz}$ and $d_{yz}$ bands is lifted (nematicity). (c) Static ARPES spectrum at 120 K collected using $s$-pol HHG light with photon energy of 29 eV.}
\label{fig1}
\end{figure}

In order to demonstrate orbital-selective electron dynamics, we studied the  polarization-dependent excitation of electrons in an archetypical Fe-based compound CaFe$_2$As$_2$, a parent Fe-pnictide superconductor exhibiting concomitant magnetic transition to the SDW ground state and structural transition to orthorhombic phase at 170 K \cite{CaFe2As2}. Angle resolved photoemission spectroscopy (ARPES)\cite{khadiza_prb18} and band structure calculations \cite{khadiza_scireports} show that the electronic structure of this material consists of three hole pockets around the Brillouin zone center, $\Gamma$ (0,0) and two electron pockets around the Brillouin zone corner ($\pi$,$\pi$).

In order to get an overview of the electronic band structure, we calculated the energy bands using the full-potential linearized augmented plane wave method as captured in the Wien2k software \cite{Wien2k}. In this self-consistent method, the convergence to the ground state was achieved by fixing the energy convergence criteria to 0.0001 Ry ($\sim$1 meV) using 10$\times$10$\times$10 $k$ points within the first Brillouin zone.
The results are shown in Fig. \ref{fig1}(a). Among the three hole bands, the inner two ($\alpha$/$\beta$ bands) possess $d_{xz}/d_{yz}$ orbital symmetry. The outer hole band, denoted by $\gamma$, has the primary contribution from the $d_{xy}$ orbital. In the SDW phase, the Fe-moments align antiferromagnetically along $x$-axis \cite{neutron_diffraction} leading to an energy gap in the $\beta$ band at the Fermi level, $E_F$ \cite{khadiza_prb18}. On the other hand, the structural transition leads to weak nematicity ($\sim$ 43 meV) as shown in Fig. \ref{fig1}(b) lifting the $d_{xz}/d_{yz}$ degeneracy \cite{nematicity}. Thus, the magnetism involves primarily $d_{xz}/d_{yz}$ electrons and $d_{xy}$ electrons are sensitive to structural changes. Here, we attempted to selectively excite electrons taking part in one kind of order leaving other electrons relatively less affected. As discussed below, we discover that this is possible by exciting the sample using light pulses with different polarizations.

Time- and angle-resolved photoemission spectroscopy (trARPES) was performed using a mode-locked Ti:sapphire laser [photon energy 1.5 eV (800 nm), pulse duration 40 fs, repetition rate 5 kHz]. The pulse was split into two parts. The major part of the intensity was used to generate high-order harmonics spanning the energy range 10-50 eV (pulse duration 40 fs) using argon medium \cite{citius}, which is used as a probe pulse. The second part of the beam was used as a pump, whose intensity was controlled with a variable attenuator based on a half wave-plate and a polarizer. We could select the desired harmonics and control their flux by means of a specially-designed grating set-up, which preserves the pulse duration. The photoemission measurements (base pressure $<$1$\times$10$^{-10}$ Torr) was performed with an R3000 analyzer and a closed cycle He-cryostat at an energy resolution of 100 meV. Single crystalline CaFe$_{2}$As$_{2}$ samples were grown by the Sn-flux method \cite{growth1,growth2}. The samples were cleaved in-situ to generate a clean and flat surface before every measurement. The dynamics of the photoexcited electrons is studied by varying the delay between the pump and  probe pulses \cite{pumpprobe,ganeshprb}.

To start with, we acquired the static (i.e., only probe, no pump) ARPES spectrum of CaFe$_2$As$_2$ along $\Gamma X$-direction. The results obtained at 120 K sample temperature using $s$-polarized photons of 29 eV are shown in Fig. \ref{fig1}(c). A qualitative comparison can be performed between the experimental spectrum (characterized by an intrinsic low energy resolution due to the use of ultra-short VUV pulses) and the simulations shown in Figs. \ref{fig1}(a) and \ref{fig1}(b): a direct inspection allows one to distinguish, in the experimental spectrum, the almost degenerate magnetic $\alpha$/$\beta$ bands from the $\gamma$ band. These data are consistent with the results obtained using synchrotron radiation \cite{khadiza_prb18}.

\begin{figure}
\includegraphics[scale=.5]{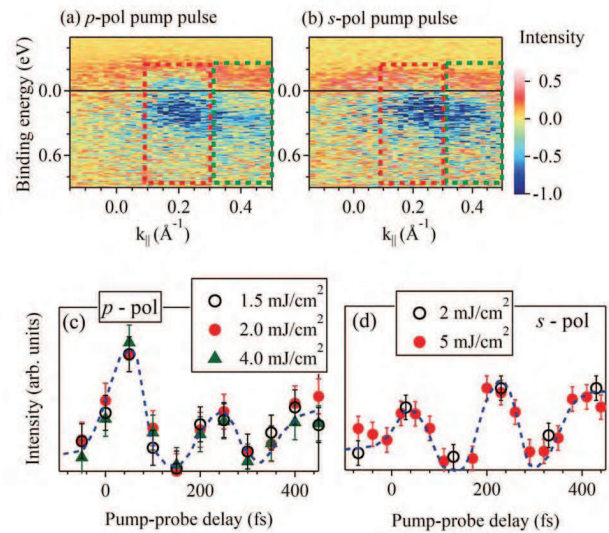}
 \vspace{-2ex}
\caption{(color online). The difference spectra for pump-probe time delay, $\Delta t$ = 200 fs with respect to the data at $\Delta t$ = -300 fs for (a) $p$-pol and (b) $s$-pol pump excitations. Sample temperature is 30 K and the probe pulse energy, $h\nu$ = 20 eV. The red box region corresponds to the ($\alpha$,$\beta$) bands and the green box region corresponds to the $\gamma$ band (see Fig. \ref{fig1}(a)). Integrated intensity between 0.1 - 0.2 eV above  $E_F$ for (c) $p$-pol and (d) $s$-pol pump pulses at different fluence as a function of pump-probe delay. Hand drawn smooth lines are guide to eye.}
\label{fig2-pol}
 \vspace{-2ex}
\end{figure}

After confirming the consistency of the static ARPES results, we carried out the pump-probe measurements to study the excited states. We used a probe energy of 20 eV, which corresponds to comparable  photoemission cross section of As 4$p$ and Fe 3$d$ states. This helps emphasizing changes of these bands having large covalency \cite{FeTe-Ganesh}, with good $k$-resolution. The pump fluence was fixed to 2 mJ/cm$^2$, which is much below the power required to drive the system into anharmonic regime \cite{phonon}. In Figs. \ref{fig2-pol}(a) and \ref{fig2-pol}(b), we show the difference of the spectra collected at 200 fs and -300 fs for (a) $p$-polarized ($p$-pol) and (b) $s$-polarized ($s$-pol) pump excitations (initial sample temperature, 30 K). This provides information on the depleted intensity in the occupied part and enhanced intensity in the unoccupied part of the electronic structure remaining at 200 fs after the pump-induced excitations. The red box region corresponds to the effect on ($\alpha$,$\beta$) bands and the green box region represents the effect on $\gamma$ band. The experimental data collected with different polarizations reveal qualitative difference!

In order to investigate if such a difference can be linked to the fluence of the pump pulse, we have carried out several measurements by varying fluence of both $p$-pol and $s$-pol light. The momentum integrated intensity in the energy range 0.1 - 0.2 eV above the Fermi level as a function of pump-probe delay are shown in Figs. \ref{fig2-pol}(c) and \ref{fig2-pol}(d) for  $p$-pol and $s$-pol excitations, respectively. The experimental results at different pump fluence exhibit identical behavior establishing absence of fluence dependence of the experimental data upto 5 mJ/cm$^2$.

To further investigate the anomaly, we define an axis system where the sample surface lies in the $xy$ plane, photoemission occurs in the $xz$-plane and the detector is placed on $z$-axis. Therefore, the electric field vector of the $p$-pol pump pulse lies in the $xz$ plane. In our setup, since the incident beam defines an angle of 36$^\circ$ with $z$-axis, the electric field vector of $p$-pol light is closely aligned with a $d_{xz}$ orbital lobe and hence, primarily excites $d_{xz}$ electrons \cite{ganeshprb}. The polarization vector of $s$-pol pulse lies in the $xy$-plane (along $y$-axis), which has a finite component along the orbital lobes of $d_{xy}$ and $d_{yz}$ states for excitations. $d_{xz}$ states is orthogonal to the polarization vector. Thus, $p$-pol pump pulse primarily excites electrons in $\alpha,\beta$ bands. On the other hand, $s$-pol light can excite electrons in all bands (including $\gamma$). This explains the difference in the spectral weights observed in Figs. \ref{fig2-pol}(a) and \ref{fig2-pol}(b).

\begin{figure}
\includegraphics[scale=.5]{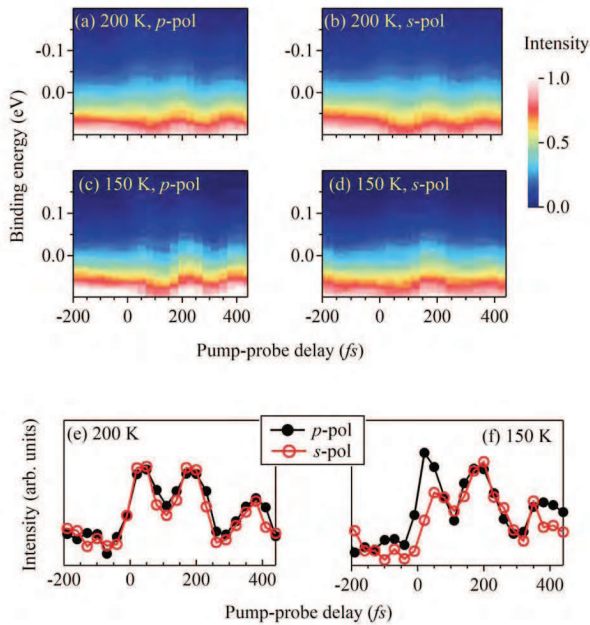}
 \vspace{-2ex}
\caption{(color online). Time-resolved spectra as a function of $\Delta t$ for (a) $p$-pol and (b) $s$-pol pump pulses with initial sample temperature of 200 K. The same at an initial sample temperature of 150 K for (c) $p$-pol and (d) $s$-pol pump pulses. The spectral intensity in the energy range 0.1 - 0.2 eV above $E_F$ at (e) 200 K and (f) 150 K. The $p$-pol (closed circles) and $s$-pol (open circles) data are superimposed to facilitate a comparative description.}
\label{fig3-pol}
 \vspace{-2ex}
\end{figure}

The excited electrons start interacting with each other and the lattice on a time scale of few tens of femtoseconds, leading to different collective excitations involving magnons, phonons, etc. Such optical perturbations may destroy the low temperature magnetic and/or structural phases. In general, the response related to electron-phonon coupling is a relatively slower process due to lattice inertia \cite{tbte3}, while the electronic phenomena like SDW order will respond very fast to the pump-induced excitation \cite{tas2}. Upon excitation, we observe that the excited electronic states get modulated by the coherent phonon mode. To analyze the response of the photoinduced perturbation, we integrated the intensity of hot electrons above $E_F$ (-0.1 to -0.2 eV). In the paramagnetic phase (initial sample temperature = 200 K), see Figs. \ref{fig3-pol}(a), \ref{fig3-pol}(b) and \ref{fig3-pol}(e), the intensity of hot electrons exhibit similar behaviour for both $p$-pol and $s$-pol excitations. However, in the SDW phase (initial sample temperature = 150 K), when the system is excited with $p$-pol light, the maximum yield of intensity occurs at $\sim$50 fs, as it is shown in Figs. \ref{fig3-pol}(c) and \ref{fig3-pol}(f); instead, a maximum is observed at about $\sim$170 fs, when pumping with $s$-pol light, see Figs. \ref{fig3-pol}(d) and \ref{fig3-pol}(f).

It is to note here that since the electron-electron (e-e) scattering time scale is too short to be detected in our measurements, the peak at 50 fs in the $p$-pol pump excitations at 150 K cannot be attributed to e-e scattering. Here, the major contribution arises due to electron-phonon and electron-magnon coupling induced effects. A fast (50 fs) shift of the leading-edge midpoint of the spectral function has been observed earlier \cite{uwe_prl2012}, which was associated to the closing of the magnetic gap. Here, employing polarized pump pulse, we are able to detect distinct signature of the magnetic excitations. The polarization dependence in the excitation of phonon modes can be attributed to the hybridization of Fe 3$d$ – As 4$p$ states, which is symmetry dependent.

The electron yield above $E_F$ can be associated, via the Fermi-Dirac distribution, to the increase of the transient electronic temperature, $T_e$, which is responsible for the melting of the magnetic and structural long-range order \cite{cr}. Thus, a peak at 50 fs time delay for the $p$-pol excitation case and its virtual absence for the $s$-pol case (intensity at 50 fs is negligibly small with the peak appearing at 170 fs) reveal an interesting scenario: the changes in magnetic and structural orders may be triggered on different time scales. Most interestingly, {\it it seems to be possible to decouple the melting processes of magnetic and structural orders}. This discovery may have relevant fundamental and applied implications for the study and manipulation of the electronic properties of various quantum materials, where it is believed that the interplay between spin, charge and lattice degrees of freedom is responsible for their exotic behaviour.

\begin{figure}
 \vspace{-4ex}
\includegraphics[scale=.4]{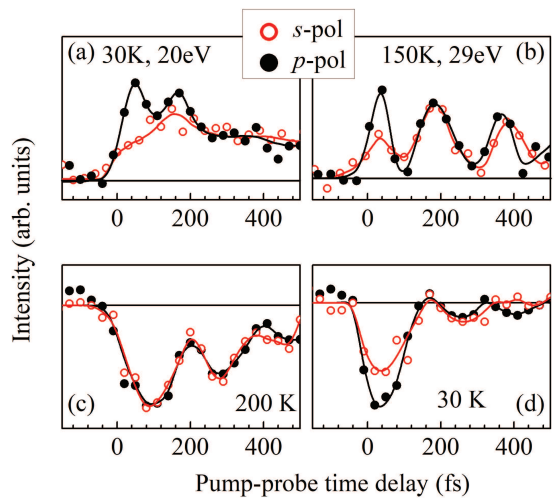}
 \vspace{-28ex}
\caption{(color online). (a) Time resolved ARPES signal (energy range 0.1 to 0.2 eV) as a function of pump-probe delay for $s$-pol and $p$-pol pump excitations with $h\nu$ = 20 eV and initial sample temperature of 30 K. (b) The same with $h\nu$ = 29 eV and initial sample temperature of 150 K. Time-resolved spectral intensity at 0.2 eV binding energy for $p$- (solid circles) and $s$- polarizations (open circles) at (c) 200 K and (d) 30 K.}
\label{fig4}
 \vspace{-2ex}
\end{figure}

In order to confirm the above scenario, we carried out some additional experiment. Figs. \ref{fig4}(a) and \ref{fig4}(b) show the trARPES spectra at 30 K for $p$- and $s$-polarizations of pump pulses with photon energy 20 eV and 29 eV, respectively; the data show the integrated photoelectron yield in the energy range 0.1 - 0.2 eV above $E_F$.  We observe that the photoelectron intensity rises very fast when the system is excited with $p$-pol light. When the system is excited with $s$-pol light, the yield of intensity rises slowly and the maximum intensity appears at a delay time of $\sim$ 200 fs. The difference observed at 150 K appears to have enhanced at 30 K.

In order to verify the transient dynamics at another $k$-point, we used probe photon energy of 29 eV, which is close to the $\Gamma$-point on the $k_z$-axis, ($k_z \approx 12\pi/c$) at an initial sample temperature of 150 K. Here, the hole pockets are relatively more separated in $k$-space than the case of $h\nu$ = 20 eV \cite{khadiza_scireports} and, hence, are expected to show better resolved behaviour of the bands of different symmetries. In addition, the relative photoemission cross-section of As 4$p$ states gets suppressed leading to dominant Fe 3$d$ contributions in the spectral intensity. Once again, the experimental results show a maximum yield of intensity at $\sim$50 fs when the system is excited with $p$-pol light, while the yield of intensity is maximum at about $\sim$200 fs when the system is excited with $s$-pol light.

In Fig. \ref{fig4}(c) and (d), we plot the $k$-integrated spectral intensity at the binding energy of 0.2 eV, which, below the critical temperature (e.g. at 30 K), reflects the yield of the folded band due to SDW order. Above threshold (200 K), we do not observe any difference between the $p$-pol and $s$-pol case, see Fig. \ref{fig4}(c). However, at 30 K, we observe significantly more intense dip in $p$-pol case compared to the $s$-pol case shown in Fig. \ref{fig4}(d). This confirms that the folded band formed due to SDW order is more efficiently depopulated in the $p$-pol case. Note that the relaxation of the hole states at 30 K is much faster compared to 200 K, which is in agreement with other measurements \cite{uwe_prl2012}.

\begin{figure}
 \vspace{-4ex}
\includegraphics[scale=.4]{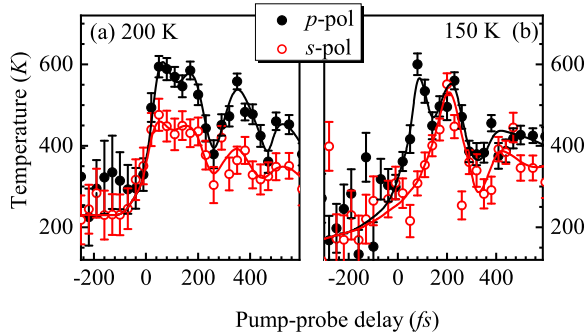}
 \vspace{-40ex}
\caption{(color online). Transient electronic temperature calculated for $p$-polarized (solid circles) and $s$ polarized (open circles) pump excitations (a) at 200 K and (b) at 150 K.}
\label{fig5}
 \vspace{-2ex}
\end{figure}

By considering the spectral function as a representation of the Fermi-Dirac distribution, we have obtained the evolution of the electron temperature after pump excitation. The experimental energy density curves (EDC) were obtained by integrating the ARPES data over a $k$-range probed along $\pm$0.3 \AA$^{-1}$. Each EDC was fitted with a Gaussian peak with an integral background multiplied by a Fermi-Dirac function, which was convoluted with the energy resolution function of the setup. The resulting electronic temperature is shown in Fig. \ref{fig5}(a) and \ref{fig5}(b) for the data at 200 K and 150 K, respectively. At 200 K, the temperature increases rapidly, soon after the pump pulse excites the electrons. Interestingly, the temperature for $p$-pol case is found to be almost 1.5 times larger than the temperature for the $s$-pol case, in the whole time-range studied indicating relatively more efficient absorption of $p$-pol light. The trend of cooling is almost identical in both the cases; this suggests that the energy transfer via electron-phonon coupling is similar for the eigenstates of different energy bands. Instead, the dynamics of the electron temperatures, obtained from the spectral functions after excitation of the sample at 150 K, are significantly different for $p$ and $s$-polarized pump pulses. The most significant difference is the appearance of a sharp peak at 50 fs in the $p$-polarized case, which is absent in the $s$-polarized case.

Here, the temperature deduced by employing the Fermi-Dirac distribution reflects the population of the unoccupied part of the electronic structure due to pump pulse excitations. Since we used same fluence for both the polarizations, there can be small difference in temperature due to differences in absorbance at difference polarizations, which will be a constant shift in temperature. Clearly, the populations of the unoccupied part, and their evolution with time, are significantly different when excited with differently polarized light, {\it both} above and below the critical temperature. At 200 K, although the trend in cooling is similar for both polarizations, indicating that energy dissipation due to electron-phonon coupling is quite similar in the two cases, $p$-polarized pump pulse is able to populate the unoccupied part more efficiently than the $s$-polarized light. In both cases, the electron temperature rises very fast soon after the pump pulse excitations. Instead, the dynamics is significantly different when the sample is initially in the SDW phase: while the $p$-polarized pumps is able to heat up the electrons at a fast time scale, the electron temperature rise is delayed in the $s$-polarized case.

In the paramagnetic phase, the $alpha$ and $beta$ bands have both $d_{xz}$ and $d_{yz}$ characters as the crystal structure is tetragonal. Therefore both $p$- and $s$-polarized pumps are able to excite them; the difference in intensity of excitations might be related to the cross section due to the component of light polarization vector along the orbital lobes. In the magnetically ordered phase, the structure becomes orthorhombic and the degeneracy of $d_{xz}$ and $d_{yz}$ electrons is lifted, leading to weak nematicity. Evidently, the $p$-polarized light is able to excite $d_{xz}$ electrons which are coupled via electron-magnon coupling and the response occurs at a fast time scale. On the other hand, $s$-polarized light is not able to excite the electrons participating in magnetic order and the slower electron temperature rise might be related to the relatively slower electron-phonon response.

It is to note here that L. Rettig {\it et al.} \cite{uwe_prl2012} have studied the perturbation dynamics for optically excited EuFe$_{2}$As$_{2}$ to investigate different coupling phenomena and reported two time scales for magnetic ordering ($\sim$50 fs) and structural orientation ($\sim$100 fs). While our results are consistent with these findings, we discover the fact that electrons corresponding to different orders can be selectively excited by using suitably polarized pump pulses. We also discover evidences that the magnetic and structural orders are decoupled, which explains for the first time, the underlying physics in the differences observed in the bulk properties. For example, Co doping in Ba(Fe$_{1-x}$Co$_{x}$)$_{2}$As$_{2}$ show that the structural transition precedes the magnetic transition \cite{co-doped bafe2as2}. On the other hand, with the application of external pressure on BaFe$_{2}$As$_{2}$, the magnetic transition precedes the structural one \cite{pressure-decoupled}. Clearly, one can play around with the structural and magnetic transition using different thermodynamic parameters.

In conclusion, we demonstrated that light polarization may be exploited to disentangle magnetic and structural excitations in quantum materials, such as Fe-based high-temperature superconductors. Employing our method, we discover that temperature-driven magnetic and structural transitions occur at different time scales; the magnetic order can be perturbed by an ultra-fast pump pulse significantly faster than the perturbation of the structural order. These results provide new insight into the exotic behaviour of iron-based superconductors. More in general, the method we have demonstrated can be applied to the study of a larger class of quantum materials as well as to their use for technological applications.

KM acknowledges financial assistance from the Department of Science and Technology, Government of India under the J.C. Bose Fellowship program and the Department of Atomic Energy under the DAE-SRC-OI Award program.




\end{document}